\documentclass[english,doublecol]{epl2}
\usepackage[T1]{fontenc}
\usepackage[latin9]{inputenc}
\usepackage{amsmath}
\usepackage{graphicx}
\usepackage{amssymb}
\usepackage{esint}

\makeatletter

\providecommand{\LyX}{L\kern-.1667em\lower.25em\hbox{Y}\kern-.125emX\@}
\providecommand{\tabularnewline}{\\}

\makeatother

\usepackage{babel}

\begin{document}

\title{Asymptotic analysis of first passage time in complex networks}
\shorttitle{Asymptotic analysis of first passage time} 

\author{
Hon Wai Lau\thanks{E-mail: \email{lau65536@gmail.com}} \and
K. Y. Szeto\thanks{E-mail: \email{phszeto@ust.hk}}
}
\shortauthor{H. W. Lau \etal}

\institute{
Department of Physics, The Hong Kong University of Science and Technology, Hong Kong, China 
} 


\abstract{
The first passage time (FPT) distribution for random walk in complex networks is calculated through an asymptotic analysis. For network with size $N$ and short relaxation time $\tau\ll N$, the computed mean first passage time (MFPT), which is inverse of the decay rate of FPT distribution, is inversely proportional to the degree of the destination. These results are verified numerically for the paradigmatic networks with excellent agreement. We show that the range of validity of the analytical results covers networks that have short relaxation time and high mean degree, which turn out to be valid to many real networks.
}

\maketitle

\global\long\def\FPT{\mathcal{F}}
\global\long\def\Prob{\mathcal{P}}
\global\long\def\PStat{\Prob^{\infty}}
\global\long\def\Walk{\mathcal{W}}
\global\long\def\Nei#1{\text{Nei}(#1)}
\global\long\def\SumNei#1#2{\sum_{#1\in\text{Nei}(#2)}}
\global\long\def\MDeg{\left\langle k\right\rangle }

Network is important in many areas of science and engineering as a
mathematical representation of the interaction of complex real systems
\cite{Albert2002-Rev,Dorogovtsev2008-Rev} such as the Internet, social
network, etc. Some of the common topological properties for many real
world networks, such as small-world and scale free properties \cite{Albert2002-Rev},
and fractal scaling \cite{Song2006-nature,Kim2007} have been investigated.
Recent research effort has begun to address the dynamical properties
and critical phenomenon in complex network. Random walk problem, due
to its simplicity, is therefore a key to the understanding of the
propagation of dynamical quantities in network. The propagation of
a random walker, which can represent the spreading of some signal
in specific application, can be measured by the time needed to reach
a selected node. In this regard, the first passage time (FPT) that
measures the first time the random walker visiting a given destination,
is critical to many events triggered by the walker, such as epidemics
spreading in social network \cite{random-walk-epidemic}, neuron firing
dynamics \cite{Intro-theo-neurobiology} and various target search
process \cite{Hughes1995-book,Redner2001-book}. Instead of the whole
distribution, the mean first passage time (MFPT) is widely studied
in different lattices with unbounded domain \cite{Redner2001-book},
and bounded domain with arbitrary shape of boundary \cite{Condamin2007a}.
Although the general solution of MFPT has been found \cite{Noh2004},
the explicit relation between first passage time and structural properties
of network is still unknown. A recent work has related MFPT to the
distance between source and destination \cite{Condamin2007-nature}
for fractal networks, which use the analysis of pseudo Green function
for finite lattice \cite{Condamin2007a}. Nonetheless, it has been
argued that this result should only be applicable to the deterministic
fractal network with homogeneous degree distribution \cite{Generalization-Fractal-Einstein-Law}.
In comparison with most deterministic fractal with fixed degree, real
world networks typically have strong inhomogeneity in degree distribution.
Thus, it suggests that the degree of nodes should play an essential
role for the random walk in complex network and a treatment differently
from the analysis of fractal network should be used.

For many real networks, the number of nodes in a given data set is
usually not big, so that a rigorous analysis on its properties is
usually difficult. Hence, schemes to adjust the size of network have
been proposed \cite{Song2006-nature,Spectal_coarse_graining}, but
it is difficult to preserve structural properties without knowing
the underlying construction mechanism. Computing the MFPT for networks
with a particular property can therefore be useful for real world
network. Our method should be complementary to the pseudo Green function
approach \cite{Condamin2007-nature}.

In this paper, we will provide an analytical expression for the asymptotic
first passage time distribution function, from which all moments,
including the mean, can be computed and compared with numerical results
for real and paradigmatic networks. We focus on the study of random
walk in complex networks with short relaxation time irrespective of
the type of network. In contrast to the pseudo Green function approach,
our asymptotic analysis can also reveal the local dynamical properties
around the destination. In particular, it allows us to compute a good
approximation of the decay rate of FPT distribution and MFPT for networks
with short relaxation time. We further argue that these expressions
are approximate bounds for other general complex networks. Since most
real world networks have short relaxation time, we find that our analytical
calculation produces excellent comparison to numerical results on
the MFPT for paradigmatic as well as real world networks.

\section{Formulation}

\begin{figure}
\begin{centering}
\includegraphics[width=0.32\textwidth]{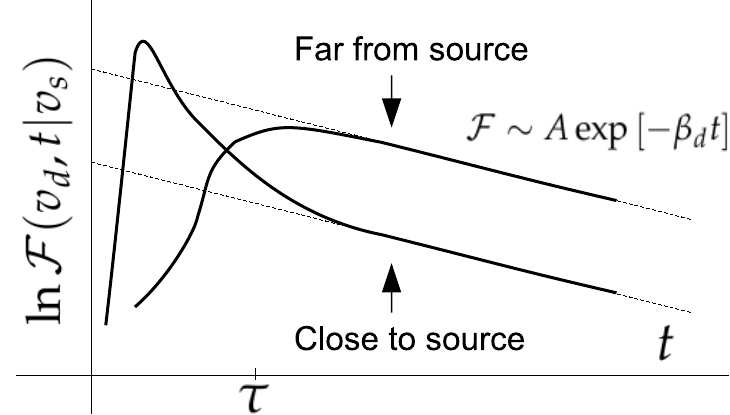}
\par\end{centering}

\caption{\label{fig:fpt-distribution} Typical FPT distribution for a destination
$v_{d}$. At the long time scale $t\gg\tau$, FPT distribution decays
exponentially. For the destination close to the source, the walker
can reach the destination in a short time, resulting in a peak higher
than the asymptotic curve. If the destination is far from the source,
the walker has to visit many other nodes before the destination, resulting
in a slow increase of FPT distribution.}

\end{figure}

We consider a finite undirected network which consists of a set of
nodes $\{v_{1,}v_{2},...,v_{N}\}$ that are connected by edges. The
set of neighbors of a node $v$ is denoted by $\text{Nei}(v)$. Let
$\Prob(v,t)$ be the probability of a random walker located at node
$v$, at time step $t$. At each time step, the walker can move from
the current node $v_{i}$ to one of its nearest neighbor node $v_{j}\in\text{Nei}(v_{i})$,
with equal probability $1/k_{i}$. Hence, the master equation of random
walk is\begin{equation}
\mathcal{P}(v_{i},t+1)=\sum_{v_{j}\in\text{Nei}(v_{i})}\frac{\mathcal{P}(v_{j},t)}{k_{j}}\label{eq:master eqt of random walk}\end{equation}
For the random walk problem in a finite domain, there is a characteristic
relaxation time $\tau$ such that equilibrium is essentially reached
when $t\gg\tau$. We denote the equilibrium probability distribution
by $\PStat(v)$ for $t\to\infty$. By the principle of detailed balance,
the net flow of walker probability $\mathcal{P}^{\infty}(v_{i})/k_{i}-\mathcal{P}^{\infty}(v_{j})/k_{j}$
is zero along each edge at equilibrium. Hence, the equilibrium probability
is proportional to the node degree $\PStat(v)\propto k_{v}$. After
normalization, we get the Kac's result \cite{Condamin2007a}, $\PStat(v)=k_{v}/N\MDeg$,
where $\MDeg$ is the mean degree of network. Therefore, at equilibrium,
the walker probability moving along each edge, in both directions,
is equal to $1/N\MDeg$.

Let $\FPT(v_{d},t|v_{s})$ be the FPT distribution from a source node
$v_{s}$ to a destination node $v_{d}$ that takes time $t$. By adding
a sink at the destination, the FPT distribution can be found by computing
the probability of the walker trapped at the sink. For a network with
sink, we denote the probability that the walker is located at node
$v$, at time $t$ by $\Walk(v,t)$. Note that $\Walk(v,t)$ is the
analog of $\Prob(v,t)$, which addresses a network without sink. The
FPT distribution for a sink at $v_{d}$ can be computed by the following
equations:\begin{equation}
\begin{cases}
\Walk(v,0)=\delta_{vv_{s}}\\
\Walk(v_{d},t)=0 & \forall t>0\\
\Walk(v,t)=\SumNei uv\frac{\Walk(u,t-1)}{k_{u}} & \forall v\in V-\{v_{d}\}\\
\FPT(v_{d},t|v_{s})=\SumNei u{v_{d}}\frac{\Walk(u,t-1)}{k_{u}}\end{cases}\label{eq:FPT-evolving}\end{equation}
The first equation is the initial condition. The second equation
is the absorbing boundary condition at the destination. The third
equation is the random walk transition probability corresponding to
(\ref{eq:master eqt of random walk}). These three equations define
the random walk problem with a sink located at $v_{d}$. The last
equation introduces the method to find the FPT distribution, which
is given by the incoming probability flux flowing towards the destination.
 Moreover, removal of the walker at the sink decreases the total
walker probability, or the survival probability $\Walk_{total}(t)$
defined by $\Walk_{total}(t)=\sum_{v\in V}\Walk(v,t)$. Hence, the
FPT distribution and the total walker probability are related by\begin{equation}
\FPT(v_{d},t|v_{s})=\Walk_{total}(t-1)-\Walk_{total}(t)\label{eq:F-W-relation}\end{equation}
as the random walker is absorbed by the destination node at time $t$.

Now, we perform asymptotic analysis of FPT distribution. When $t\gg\tau$,
the initial information of the source node is washed away, so walker
probability decreases uniformly for each node in the network. In this
case, $\Walk(v,t)$ can be separated into two parts: \begin{equation}
\Walk(v,t)\sim\Walk(v)\Walk_{total}(t),\qquad t\gg\tau\label{eq:Walk-equilibrium assumption}\end{equation}
where $\Walk(v)$ is a time independent probability distribution that
depends on network topology and the location of destination. By substituting
Eq. (\ref{eq:Walk-equilibrium assumption}) back into the last equation
in (\ref{eq:FPT-evolving}), we can get a structural factor $\beta_{d}$
that depends on the destination node $v_{d}$:\begin{equation}
\beta_{d}=\SumNei u{v_{d}}\frac{\Walk(u)}{k_{u}}\label{eq:FPT-decay rate}\end{equation}
and the asymptotic form of FPT distribution $\FPT(v_{d},t|v_{s})\sim\beta_{d}\Walk_{total}(t-1)$
for $t\gg\tau$. Solving this equation along with Eq. (\ref{eq:F-W-relation}),
we have $\Walk_{total}(t)\sim(1-\beta_{d})\Walk_{total}(t-1)$ and
so $\Walk_{total}(t)\propto(1-\beta_{d})^{t}$ asymptotically. As
we will show later, $\beta_{d}$ is small, so that $\Walk_{total}(t)\propto\exp\left[-\beta_{d}t\right]$.
With the known asymptotic form of $\Walk_{total}(t)$, we can conclude
that the FPT distribution is $\FPT(v_{d},t|v_{s})\sim A\exp\left[-\beta_{d}t\right]$.
Thus, FPT distribution has an exponential tail with decay rate $\beta_{d}$
as shown in Fig. \ref{fig:fpt-distribution}. In addition, we now
know a method to compute decay rate analytically by Eq. (\ref{eq:FPT-decay rate}),
provided that $\Walk(u)$ of all neighbors of destination $v_{d}$
is known.

\section{Decay rate of FPT distribution}

\begin{figure}
\begin{centering}
\includegraphics[width=0.49\textwidth]{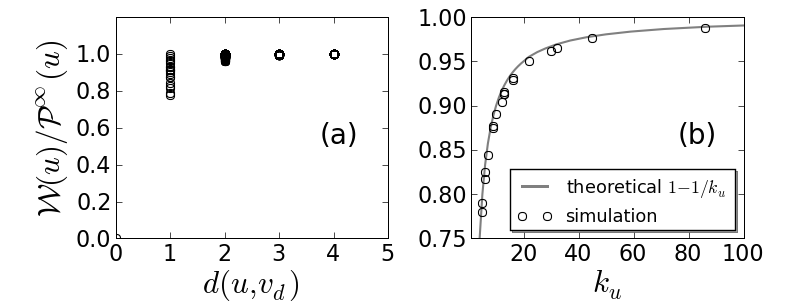}
\par\end{centering}

\caption{\label{fig:W(u)/Weq(u) - ba} For $t\gg\tau$, we show the deviation
$\Walk(u)/\PStat(u)$ of the time independent probability distribution
$\Walk(u)$ from the equilibrium distribution $\PStat(u)$ when there
is no sink, as a function of (a) distance from destination $v_{d}$,
(b) degree of nearest neighbors $u\in\Nei{v_{d}}$ which correspond
to the distance=1 in (a). The network used is the BA model with $N=10000$
and $\MDeg=10$.}

\end{figure}

After deriving the exponential tail of the FPT distribution, our next
task is to compute the decay rate $\beta_{d}$ for the network with
short relaxation time $\tau$ such that $\tau\ll1/\beta_{d}$. Physically,
$\tau$ represents the time to reach equilibrium in a network without
sink, and $1/\beta_{d}$ is the approximate time scale for the decrease
in total walker probability of the network when there is a sink to
absorb the random walker.  For network with short relaxation time,
the removal of walker probability around the destination can be accomplished
rather quickly. On the other hand, the total probability decreases
with a much slower rate and this result in a quasi-equilibrium state.
This physical picture implies that the quantity $\Walk(u)=\lim_{t\gg\tau}\Walk(u,t)/\Walk_{total}(t)$
is approximately equal to $\Prob^{\infty}(u)$. This can be observed
for the  Erdos-Renyi (ER) network and Barabasi-Albert (BA) network
which have very short relaxation time (see Table \ref{tab:rw-properties}).
One of them is illustrated numerically in Fig. \ref{fig:W(u)/Weq(u) - ba}a.
In the simplest case, it is reasonable to assume that $\Walk(u)=\PStat(u)$
for $d(u,v_{d})\ge1$. While this assumption is substituted back to
Eq. (\ref{eq:FPT-decay rate}), the decay rate is $\beta_{d}=k_{d}/N\MDeg$.
It scales linear with the degree of destination and scales inversely
with the size of the network. Thus, for a large network, $\beta_{d}$
can be very small.

Next, we proceed to a better approximation for the dependence of $\Walk(u)$
on the $k_{u}$ around the destination. For a nearest neighbor $u$
of the destination, $\Walk(u)$ has to take a smaller value than $\PStat(u)$
to compensate the flow of walker toward the sink. Hence, it is natural
to expect that $\Walk(u)\le\PStat(u)$. Rather than assuming $\Walk(u)=\PStat(u)$
for nodes other than destination, a refined assumption can be made
for the next nearest neighbors:\begin{equation}
\Walk(u,t)=\PStat(u)\Walk_{total}(t),\qquad d(u,v_{d})\ge2\label{eq:Walk-assumption d.ge.2}\end{equation}
With this assumption, $\Walk(u)$ for $u\in\Nei{v_{d}}$ has to be
found before computing the decay rate. Here, nodes with distance two
from the destination are treated as the reservoir of walker probability.
We thus assume that each edge of node $v$, that are of distance greater
than 1 away from the destination, $d(v,v_{d})\ge2$, has probability
$1/N\MDeg$ moving out at each time step.

Now, let us focus on a nearest neighbor node $u$ of the destination.
Since there is no probability flowing from $v_{d}$ to $u$, therefore,
only $k_{u}-1$ neighbors of $u$ have probability flowing into $u$.
If there are no edges connecting between neighbors of $v_{d}$, i.e.,
assuming zero clustering coefficient of $v_{d}$, then each edge contributes
$1/N\MDeg$ to $u$ from the second neighbor of $v_{d}$, or $\Walk(u)=\left(k_{u}-1\right)/N\MDeg$.
If there are some edges connecting neighbors of $v_{d}$, i.e., for
the case of non-zero clustering coefficient, then some neighbors of
$u$ are also neighbors of $v_{d}$. In this case, nodes in $\Nei{v_{d}}$
have value less than $\Prob^{\infty}$ and the probability moving
out has value less than or equal to $1/N\MDeg$. The exact value can
be found by solving $\Walk(u)$ for all $u\in\Nei{v_{d}}$ simultaneously
and it is upper bounded by\begin{equation}
\Walk(u)\le\left(\frac{k_{u}-1}{k_{u}}\right)\PStat(u),\qquad d(u,v_{d})=1\label{eq:rwnet-W(v) of nearest nei}\end{equation}
As shown in Fig. \ref{fig:W(u)/Weq(u) - ba}b for the BA model, this
result fits well with the simulation of $\Walk(u)$ and the equality
sign holds approximately. In general, this upper bound can be reached
for networks with very short relaxation time. Now we put Eq. (\ref{eq:rwnet-W(v) of nearest nei})
into Eq. (\ref{eq:FPT-decay rate}) to get\begin{equation}
\beta_{d}\le\frac{k_{d}}{N\MDeg}\left(1-\frac{1}{k_{d}}\SumNei u{v_{d}}\frac{1}{k_{u}}\right)\label{eq:FPT-decay rate-2-pre}\end{equation}
where the last term is the mean of inverse degree of neighbor that
can be approximated by mean field. Let's recall that the probability
of selecting one of nearest neighbors with degree $k$ for uncorrelated
network is $P(k)k/\MDeg$ \cite{Newman2001}, where $P(k)$ is the
degree distribution of the network. Similar to the computation of
mean degree of neighbor in Ref. \cite{Ma&Szeto2006}, the mean of
inverse degree of neighbor is $\int_{0}^{\infty}\left(P(k)k/\MDeg\right)\frac{1}{k}dk=\frac{1}{\MDeg}$,
so \begin{equation}
\beta_{d}\le\frac{k_{d}}{N\MDeg}\left(1-\frac{1}{\MDeg}\right)\label{eq:FPT-decay rate-2}\end{equation}
which suggests that the actual decay rate obtained in simulation is
slower than this upper bound. Note that this result is consistent
with our basic hypothesis $\tau\ll1/\beta_{d}$ if $\tau\ll N$. We
find that this theoretical decay rate is in excellent agreement with
the simulation result for the ER and BA network (see Table \ref{tab:rw-properties}).

\section{Mean first passage time}

\begin{figure}
\begin{centering}
\includegraphics[width=0.49\textwidth]{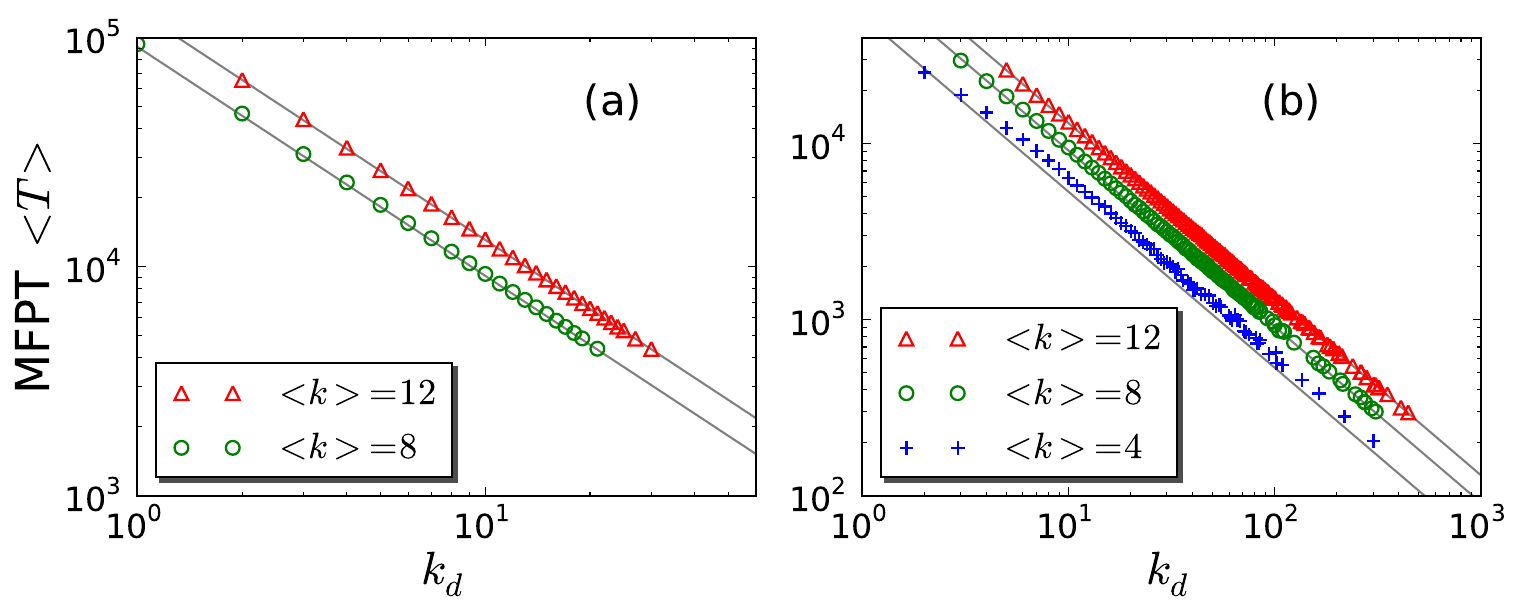}
\par\end{centering}

\caption{\label{fig:mfpt-vs-k for ER and BA} (color online) MFPT vs degree
of destination $k_{d}$ for the (a) ER and (b) BA model, with $N=10000$.
The points are simulation result and the lines are the theoretical
prediction.}

\end{figure}

Since FPT distribution is normalized for a finite undirected network,
$\int_{0}^{\infty}\FPT(v_{d},t|v_{s})dt=1$, it can be split asymptotically
as the sum of short time scale and long time scale as:\begin{equation}
\int_{0}^{c}\FPT(v_{d},t|v_{s})dt+\int_{c}^{\infty}Ae^{-\beta_{d}t}dt\sim1,\qquad c\gg\tau\label{eq:FPT-asymptotic}\end{equation}
where $c$ is a cutoff value. In a finite domain, FPT distribution
always rise from 0 at $t=0$ and decreases with exponential tail \cite{Redner2001-book}
when $t\to\infty$ (see Fig. \ref{fig:fpt-distribution}). For a destination
node far from the source node, $\FPT(v_{d},t|v_{s})$ is bounded above
by $Ae^{-\beta_{d}t}$ within $0<t<c$. Hence, for short relaxation
time $\tau\ll c\ll1/\beta_{d}$, the first term is bounded by $\int_{0}^{c}Ae^{-\beta_{d}t}dt\sim0$
which contributes negligibly small value to the sum. By extending
the domain from $c$ to 0 for the dominant second term, we obtain
$A\sim\beta_{d}$ by computing $\int_{0}^{\infty}Ae^{-\beta_{d}t}dt\sim1$.
Similar analysis can be performed for MFPT $\left\langle T\right\rangle =\int_{0}^{\infty}t\FPT(v_{d},t|v_{s})dt$
and the result is $\left\langle T\right\rangle \sim1/\beta_{d}$.
Using the decay rate derived in Eq. (\ref{eq:FPT-decay rate-2}),
the MFPT is given by\begin{equation}
\left\langle T(v_{d})\right\rangle \ge\frac{N\MDeg}{k_{d}}\left(\frac{1}{1-\MDeg^{-1}}\right)\label{eq:MFPT-2}\end{equation}
which is a lower bound for the MFPT. We can get a similar approximate
result for $\left\langle T(k_{d})\right\rangle $ of MFPT averaged
over all node pairs with same destination degree because this average
is dominated by nodes separated far away. Networks with short relaxation
time $\tau\ll N$ such as the ER and BA model have diameter of order
$\mathcal{O}(\ln N)$ and the number of nodes grow exponential as
the distance increase. So the MFPT is dominated by the nodes with
separation of order $\mathcal{O}(\ln N)$. We see excellent agreement
of our theory with numerical results for the ER and BA networks in
Fig. \ref{fig:mfpt-vs-k for ER and BA}. However, for networks with
fractal scaling \cite{Song2006-nature}, the diameter scales as $\mathcal{O}(N^{\alpha})$
for some $\alpha$, so that more details about the structure are required
to compute MFPT. Hence, we do not expect our result is good when applied
to those fractal networks. Note that Eq. (\ref{eq:MFPT-2}) is compatible
with a previous numerical result \cite{Baronchelli2006} suggesting
that $\left\langle T(k_{d})\right\rangle \propto1/k_{d}$, in which
we have also computed the explicit form and shown that it is the lower
bound of MFPT of networks with short relaxation time. With the method
we used, higher moments of FPT distribution can also be computed as
$\left\langle T^{n}(k_{d})\right\rangle \approx n!\beta_{d}^{-n}$.
Note also that our derivation depends on the property of short relaxation
time regardless of the size of network. In addition, the mean first
passage time $\left\langle T\right\rangle _{G}$ of the whole network
$G$ can be computed by taking weighted average over the degree distribution
as $\left\langle T\right\rangle _{G}=\sum P(k)\left\langle T(k)\right\rangle $.
Since $P(k)\to0$ as $k\to\infty$ and $\left\langle T(k)\right\rangle \propto1/k$
for the networks considered in this paper, we have $\left\langle T\right\rangle _{G}\propto N\left\langle k\right\rangle $
which grows linear in the network size.

\section{Real world networks}

\begin{table*}
\caption{\label{tab:rw-properties} Random walk properties of different paradigmatic
and real world networks, with size $N$, mean degree $\left\langle k\right\rangle $,
relaxation time $\tau$, simulation decay rate $\beta_{d}^{*}$ and
theoretical decay rate $\beta_{d}$ given by Eq. (\ref{eq:FPT-decay rate-2}).
The method to compute $\tau$ is described in text and the $\beta_{d}^{*}$
can be found numerically after a long run by using Eq. (\ref{eq:FPT-decay rate}).}

\centering{}\begin{tabular}{lccccc}
\hline 
Networks & $N$ & $\left\langle k\right\rangle $ & $\tau$ & $\frac{1}{\left\langle \beta_{d}^{*}\right\rangle }$ & $\left\langle \frac{\beta_{d}^{*}}{\beta_{d}}\right\rangle $\tabularnewline
\hline
ER model & 10000 & 12 & 1.7 & 10996 & 0.99\tabularnewline
BA model & 10000 & 12 & 1.6 & 10985 & 1.00\tabularnewline
BA model & 10000 & 8 & 2.3 & 11624 & 0.99\tabularnewline
BA model & 10000 & 4 & 5.6 & 14457 & 0.98\tabularnewline
Coauthorship network, astro-ph \cite{Newman2001a,Newman2001b} & 14845 & 16.1 & 304 & 18435 & 0.82\tabularnewline
C. Elegans neural network \cite{CElegans-neural-network} & 297 & 14.5 & 4.6 & 323 & 0.99\tabularnewline
Coauthorship network, cond-mat \cite{Newman2001a,Newman2001b} & 13861 & 6.4 & 139 & 23601 & 0.70\tabularnewline
E. Coli. Metabolic network \cite{metabolic-network} & 2268 & 5.0 & (16) & 3190 & 0.92\tabularnewline
Coauthorship network, hep-th \cite{Newman2001a,Newman2001b} & 5835 & 4.7 & 179 & 10955 & 0.69\tabularnewline
Yeast protein interaction \cite{Protein-interaction-network-yeast} & 1458 & 2.7 & 120 & 3398 & 0.83\tabularnewline
Western State Power Grid \cite{Watts1998-nature} & 4941 & 2.7 & 3689 & 30008 & 0.33\tabularnewline
\hline
\end{tabular}
\end{table*}
\begin{figure}
\begin{centering}
\includegraphics[width=0.49\textwidth]{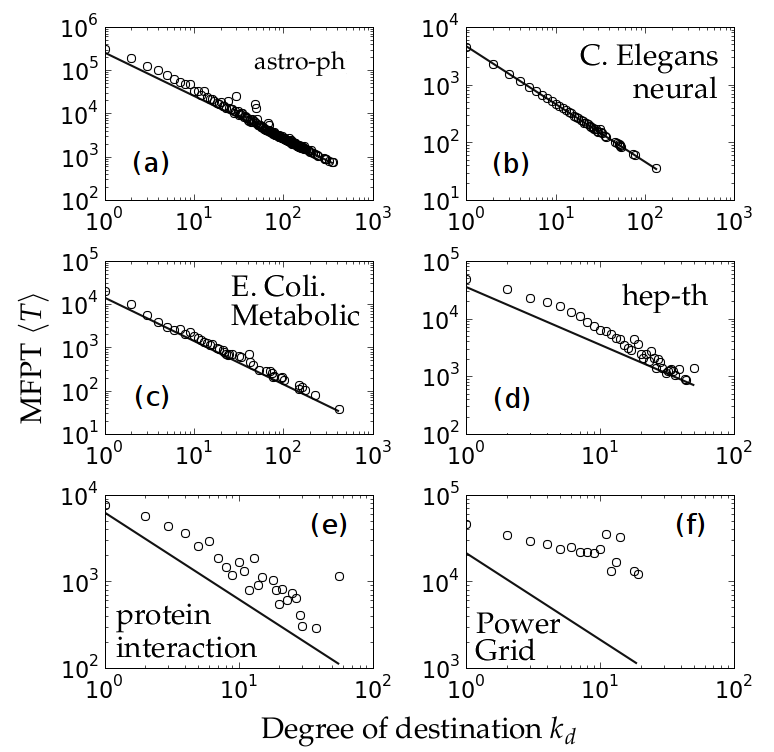}
\par\end{centering}

\caption{\label{fig:MFPT-real-networks-1} MFPT vs degree of destination $k_{d}$
for real world networks. The circles represent simulation result and
the curves represent computed MFPT.}

\end{figure}

We now test the applicability of our theory to real world networks
which may have long relaxation time and structural inhomogeneity.
The networks we examined are the Coauthorship network (astro-ph, cond-mat,
hep-th) \cite{Newman2001a}, C. Elegans neural network \cite{CElegans-neural-network},
E. Coli. Metabolic network \cite{metabolic-network}, Yeast protein
interaction \cite{Protein-interaction-network-yeast} and Western
State Power Grid \cite{Watts1998-nature}. In the simulation, only
the largest connected component of each undirected network is considered.

The random walk properties and MFPT for different networks are presented
in Table \ref{tab:rw-properties} and Fig. \ref{fig:MFPT-real-networks-1}.
The relaxation time $\tau$ is computed by $\tau=-1/\ln(|\lambda_{2}|)$,
where $\lambda_{2}$ is the second largest eigenvalue%
\footnote{E. Coli. metabolic network is a bipartite network that has no odd
loops, so its second largest eigenvalue is -1 which corresponding
to a state oscillating between two different partitions. Nevertheless,
the equilibrium for odd and even time series can be reached separately
and the first passage time is still well defined. Instead, the third
eigenvalue $\lambda_{3}$ can be used to represent the approximate
relaxation time to reach the oscillating state.%
} of the time evolution operator Eq. (\ref{eq:master eqt of random walk}),
or the random walk matrix (see \cite{Spectral_densities_of_sf}).
From Table \ref{tab:rw-properties}, we can draw the following observations.
(1) The decay rates are bounded by our theoretical value for all networks
tested ($\left\langle \beta_{d}^{*}/\beta_{d}\right\rangle $ is always
less than 1). This observation strongly suggests that our result of
decay rate can be applied to real world networks. (2) The inverse
of the decay rate is approximated equal to the network size for networks
with short relaxation time, in agreement with Eq. (\ref{eq:FPT-decay rate-2})
(3) For BA and ER networks, the relaxation time $\tau$ is short,
of the order $\mathcal{O}(\ln N)$. In fact, the relaxation time of
a BA network is shorter than the corresponding ER network with the
same size and mean degree, which implies that the propagation in the
BA network is faster than the ER network. This is caused by the shorter
diameter of the BA network $\mathcal{O}(\ln N/\ln\ln N)$, while for
ER it is of order $\mathcal{O}(\ln N)$ \cite{sf-network-ultrasmall}.
Our theory of FPT distribution works very well for the ER and BA networks,
as they both have short relaxation times. (4) For those real world
networks with relaxation time $\tau$ and the inverse of decay rate
$1/\left\langle \beta_{d}^{*}\right\rangle $ smaller by two orders
of magnitude, we obtain good agreements between theory and numerical
results. For example, the C. Elegan neural network has the shortest
relaxation time of order $\mathcal{O}(\ln N)$ in which the decay
rate and MFPT agree very well in both Table \ref{tab:rw-properties}
and Fig. \ref{fig:MFPT-real-networks-1}b. For other networks, there
are some deviations. The result on C. Elegan suggests that its neural
network has a very efficient structure that allows a very fast transmission
of signal between different neurons. This kind of insight is a valuable
spin-off of our theoretical analysis.

Returning to the computation of relaxation time, we need to consider
global information of the network. In general, we expect the relaxation
time will decrease with increasing mean degree of the network, as
shown by BA networks in Table \ref{tab:rw-properties}. For many real
world networks with high mean degree, their relaxation times are also
short as they are similar to our examples of random paradigmatic networks.
In contrast, for network with low mean degree, say $\MDeg\approx2$,
they are better described by trees, so that the walker needs more
time to move between nodes as their separation is generally larger,
implying a longer relaxation time. For these networks of low mean
degree, deviation from our theoretical bound is expected. The above
discussions on the relation between the dynamical property in terms
of relaxation time and static property in terms of mean degree can
be observed in Fig. \ref{fig:MFPT-real-networks-1}. Specifically,
for the two similar coauthorship networks in Fig. \ref{fig:MFPT-real-networks-1}a
and \ref{fig:MFPT-real-networks-1}d, the network with smaller mean
degree has larger deviation. On the other hand, both the protein interaction
network and the power grid have same mean degree, but the protein
interaction network has shorter relaxation time and fits better by
our theory. The long relaxation time of the power grid, a consequence
of its two dimensional planar nature with long diameter \cite{Watts1998-nature},
shows that our result is not a good fit, but still provides a lower
bound. For the power grid, our assumption about equilibrium states
of second nearest neighbor fails as it has long relaxation time and
strong local connection. Our theoretical prediction overestimates
the local $\Walk(u)$ and the decay rate. The actual decay rate in
this power grid is less correlated with the degree of the node, which
implies that the MFPT for the high degree nodes is similar to the
low degree nodes, resulting in a relatively flat distribution in Fig.
\ref{fig:MFPT-real-networks-1}f. For networks similar to the power
grid, first passage time distribution is better described by the distance
between source and destination.

Finally, we have also observed a correlation between the random walk
centrality \cite{Noh2004} and the measured decay rate. In fact, our
numerical results show that the random walk centrality is approximately
equal to the measured decay rate. This will be investigated further
in future research.

\section{Summary and discussion }

In conclusion, our asymptotic analysis of FPT distribution allows
us to compute an accurate expression of the decay rate of FPT distribution
(\ref{eq:FPT-decay rate-2}) for networks with short relaxation time
$\tau\ll1/\beta_{d}$. We have also shown that the decay rate of real
world networks has similar order of magnitude as $\PStat(v_{d})=k_{d}/N\MDeg$
and is upper bounded by this value. This upper bound can be explained
by a physical picture that local neighbors around the destination,
at the quasi-equilibrium state, should take the walker probability
less than $\PStat(u)$. Moreover, we have computed the MFPT (\ref{eq:MFPT-2})
for destination far from the source. A similar expression of MFPT
that depends on destination degree is argued to be correct for network
with $\tau\ll N$. These theoretical results on the relationship between
MFPT and destination degree have been verified by numerical simulations
on both paradigmatic and real world networks. Our analysis can also
be readily extended to other variants of random walk problem, such
as random walk in weighted networks and biased random walk \cite{bias_rw_local},
with short relaxation time and it predicts the MFPT to be \begin{equation}
\left\langle T(k)\right\rangle \gtrsim\frac{1}{\PStat(k)}\end{equation}
where $\PStat(k)$ is the equilibrium probability for the corresponding
random walk problem.

Since $1/\beta_{d}$ also scales linearly in $N$, the condition for
networks with short relaxation time then becomes $\tau\ll N$. Assuming
that this condition of short relaxation time is met, we have shown
numerically that the relaxation times of many real world networks
are actually short. Therefore, our analytical results are actually
applicable to these real networks. We also argued that real networks
with high mean degree should have short relaxation time.

Finally, we should note that although paradigmatic networks with random
global links, such as the ER and BA network, show extremely short
relaxation time \cite{Samukhin2008} and our theory fits very well
with the numerical data on these networks, we expect more complex
dynamical properties from real networks, which deserve further study.

\acknowledgments

K. Y. Szeto acknowledges the support of grant CERG 602506 and 602507.
We also thank the discussion with Shi Qinwei.

\bibliographystyle{eplbib}
\bibliography{ref}

\end{document}